\documentclass{sigchi-ext}
\usepackage[T1]{fontenc}
\usepackage{textcomp}
\usepackage[scaled=.92]{helvet} % for proper fonts
\usepackage{graphicx} % for EPS use the graphics package instead
\usepackage{balance}  % for useful for balancing the last columns
\usepackage{booktabs} % for pretty table rules
\usepackage{ccicons}  % for Creative Commons citation icons
\usepackage{ragged2e} % for tighter hyphenation
\usepackage[subtle]{savetrees}

\newcommand{\systemname}{{UbiNIRS}}

\newcommand{\etal}{\textit{et al}. }
\newcommand{\ie}{\textit{i}.\textit{e}.}
\newcommand{\eg}{\textit{e}.\textit{g}.}

\def\plaintitle{SIGCHI Extended Abstracts Sample File: Note Initial
  Caps} 
\def\emptyauthor{}
\def\plainkeywords{NIRS; software framework; mobile sensing.}

\title{UbiNIRS: A Software Framework for Miniaturized NIRS-based Applications}

\numberofauthors{6}
\author{%
  \alignauthor{%
    \textbf{Weiwei Jiang $^1$} \\
    \email{weiwei.jiang\\@student.unimelb.edu.au} \\ 
    \textbf{Zhanna Sarsenbayeva $^1$} \\ % 1
    \email{zhanna.sarsenbayeva\\@unimelb.edu.au} \\
    \textbf{Difeng Yu $^1$} \\
    \email{difeng.yu\\@student.unimelb.edu.au}\\
    \textbf{Jing Wei $^1$} \\
    \email{jing.wei\\@student.unimelb.edu.au}\\
    }\alignauthor{%
    \textbf{Niels van Berkel $^2$}\\ % 2
    \email{nielsvanberkel@cs.aau.dk}
    \textbf{Jorge Goncalves $^1$}\\ % 4
    \email{jorge.goncalves@unimelb.edu.au}
    \textbf{Vassilis Kostakos $^1$}\\ % 5
    \email{vassilis.kostakos@unimelb.edu.au}} \\ \newline \vspace{2em} \\
    \affaddr{$^1$ The University of Melbourne,} 
    \affaddr{Parkville VIC 3010, Australia} \\
    \affaddr{$^2$ Aalborg University,} 
    \affaddr{Fredrik Bajers Vej 5, 9100 Aalborg, Denmark} \vfil 
}

\definecolor{linkColor}{RGB}{6,125,233}
\hypersetup{%
  pdftitle={\plaintitle},
  pdfauthor={\emptyauthor},
  pdfkeywords={\plainkeywords},
  bookmarksnumbered,
  pdfstartview={FitH},
  colorlinks,
  citecolor=black,
  filecolor=black,
  linkcolor=black,
  urlcolor=linkColor,
  breaklinks=true,
}

\begin{document}

%% For the camera ready, use the commands provided by the ACM in the Permission Release Form.
% \CopyrightYear{2020}
% \setcopyright{rightsretained}
\setcopyright{none}
% \conferenceinfo{CHI'20,}{April  25--30, 2020, Honolulu, HI, USA}
% \isbn{978-1-4503-6819-3/20/04}
% \doi{https://doi.org/10.1145/3334480.XXXXXXX}
%% Then override the default copyright message with the \acmcopyright command.
% \copyrightinfo{\acmcopyright}
\copyrightinfo{}

\maketitle

% Uncomment to disable hyphenation (not recommended)
% https://twitter.com/anjirokhan/status/546046683331973120
\RaggedRight{} 

% Do not change the page size or page settings.
\begin{abstract}
%Recent work from the HCI community shows great potential for the use of miniaturized near-infrared spectroscopy (NIRS), an emerging technology. NIRS can identify objects by sensing their ingredients rather than their appearances, resulting in several advantages in comparison to vision-based classification methods. However, developing such an application is non-trivial as it requires significant programming efforts and professional knowledge of NIRS. 
We present {\systemname}, a software framework for rapid development and deployment of applications using miniaturized near-infrared spectroscopy (NIRS). NIRS is an emerging material sensing technology that has shown a great potential in recent work from the HCI community such as \textit{in situ} pill testing. However, existing methods require significant programming efforts and professional knowledge of NIRS, and hence, challenge the creation of new NIRS-based applications. Our system helps to resolve this issue by providing a generic server and a mobile app, using the best practices for NIRS applications in literature. The server creates and manages {\systemname} instances without the need for any coding or professional knowledge of NIRS. The mobile app can register multiple {\systemname} instances by communicating with the server for different NIRS-based applications. Furthermore, {\systemname} enables NIRS spectrum crowdsourcing for building a knowledge-base. %-- a typically time-consuming and costly process.
\end{abstract}


% ACM classification.



% Print the classficiation codes
%Please use the 2012 Classifiers and see this link to embed them in the text: \url{https://dl.acm.org/ccs/ccs_flat.cfm}

\section{Introduction}
Despite the widespread availability of advanced sensors, there is still no feasible way to obtain information on the ingredients of objects in our daily life. Conventional methods require samples to be sent to a dedicated laboratory for sophisticated analysis. However, recently emerging miniaturized near-infrared spectroscopy (NIRS) enables the analysis of an object's ingredients \textit{in situ}~\cite{nirsystems2002guide, siesler2008near}. NIRS utilizes near-infrared lights in multiple wavelengths between $780~nm$ and $2500~nm$. This approach leverages the phenomenon that different ingredients (molecules) absorb different wavelengths of near-infrared lights. By emitting near-infrared lights to an object and measuring the intensities reflected back from the object in those wavelengths, a near-infrared spectrum can be acquired. This spectrum can be regarded as a unique chemical ``fingerprint'' for the object, from which ingredient information can be extracted. 

\begin{marginfigure}[-20pc]
  \begin{minipage}{\marginparwidth}
    \centering
    \includegraphics[width=0.9\marginparwidth]{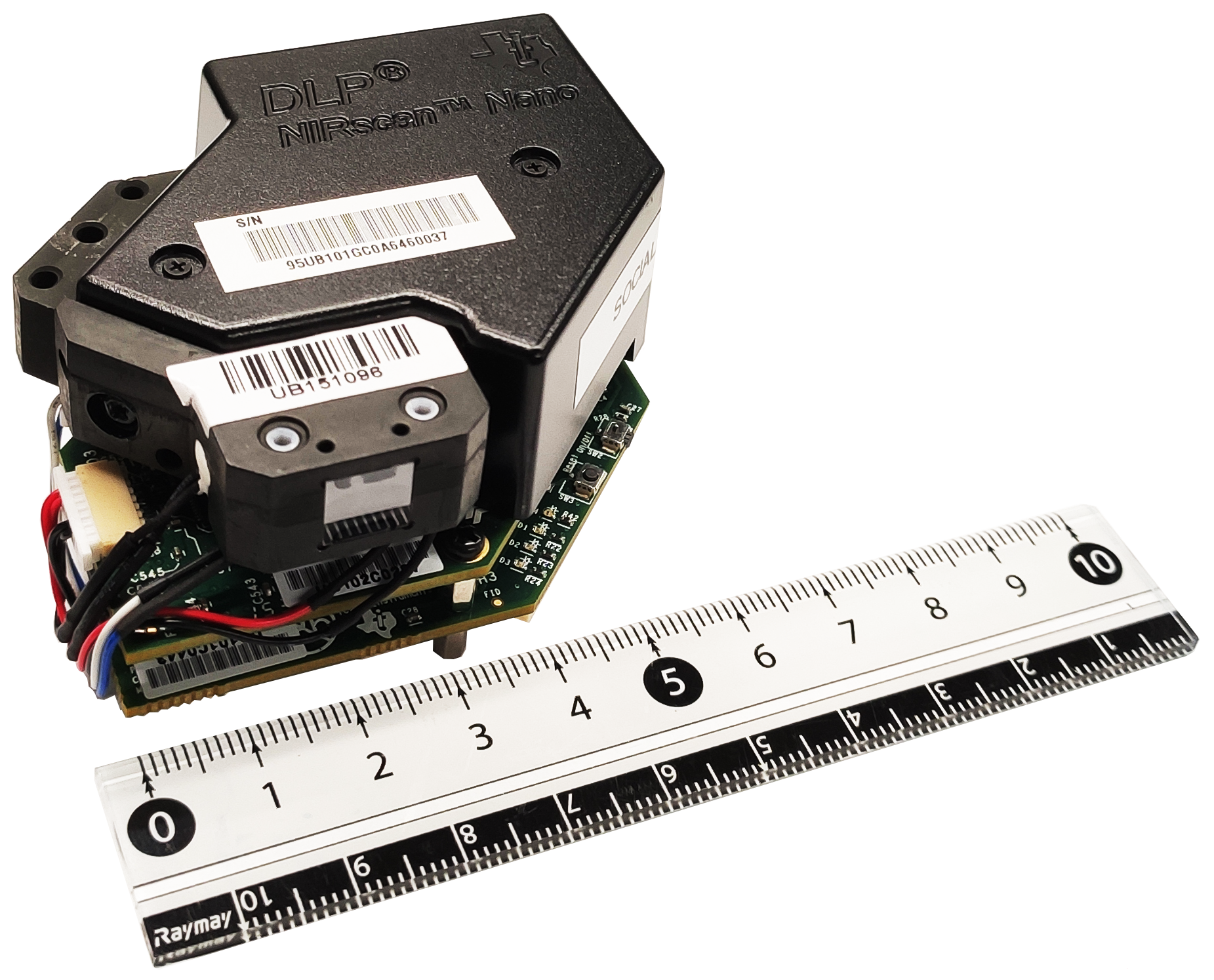}
    \caption{A miniaturized NIRS \protect\cite{dlpnirscanonline}.}
    \label{fig:nirs_scanner}
  \end{minipage}
\end{marginfigure}

\begin{marginfigure}[-3pc]
  \begin{minipage}{\marginparwidth}
    \centering
    \includegraphics[width=0.9\marginparwidth]{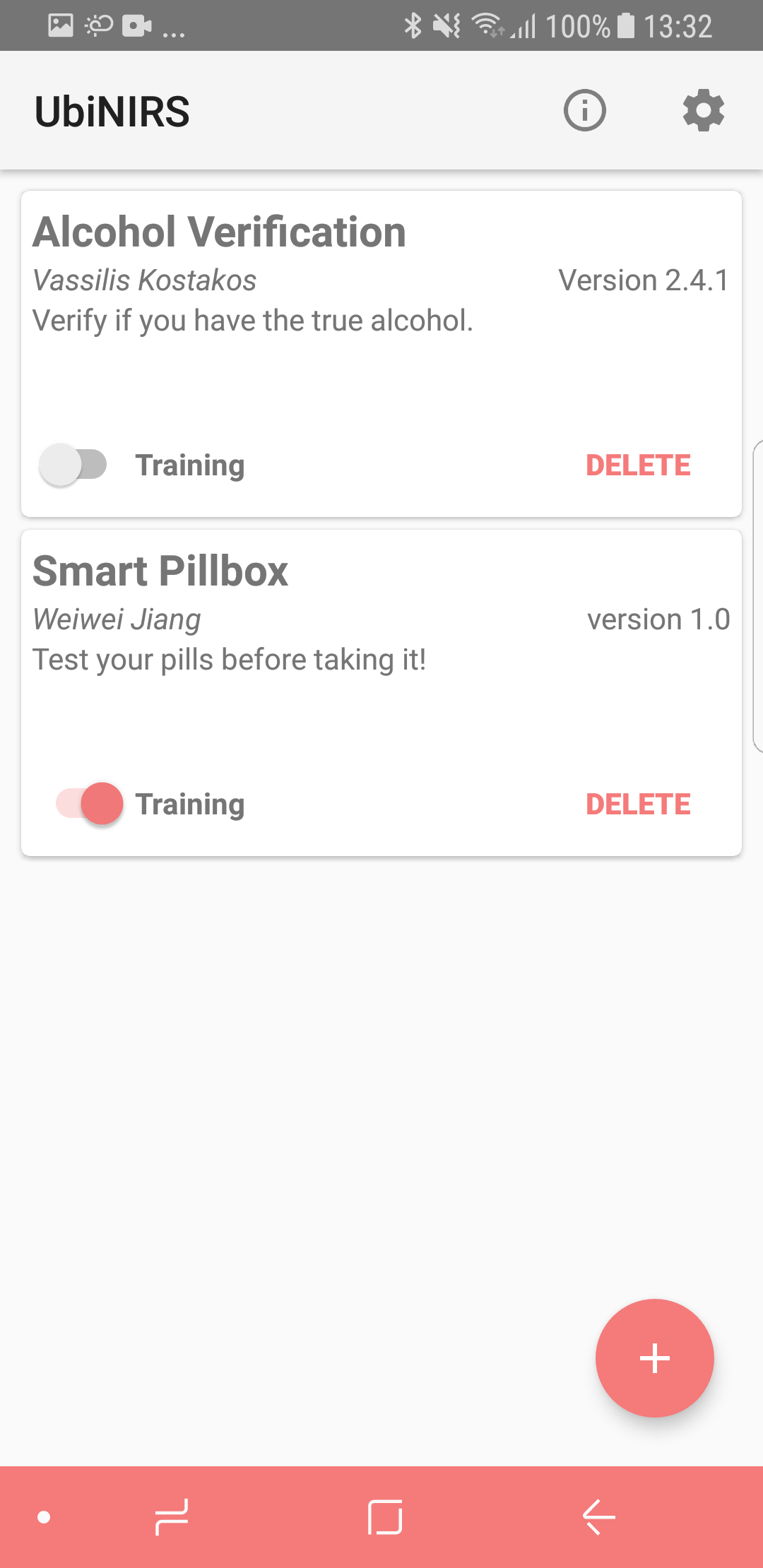}
    \caption{UbiNIRS mobile app.}
    \label{fig:app_main}
  \end{minipage}
\end{marginfigure}

However, recent studies in the HCI community suggest that it is not straightforward to use the miniaturized NIRS technology in practice~\cite{jiang2018mobile, jiang2019probing, klakegg2016}. For example, Klakegg~\etal highlighted various challenges for non-experts to take advantage of miniaturized NIRS, including spectrum distortions as the result of user-induced errors~\cite{ klakegg2017towards, klakegg2016}. Nevertheless, a follow-up study by Klakegg \etal demonstrated that it was possible to assist non-experts in identifying medical pills more accurately through the use of a well-designed prototype and usable software~\cite{klakegg2018assisted}. Their study can be generalized to solid objects, such as estimating the maturity of fruits, ranking beef, etc. For the identification of non-solid objects, Jiang \etal extended the study to identify liquids such as everyday drinks and liquors using a customized 3D printed clamp and software~\cite{jiang2018mobile, jiang2019probing}.

The aforementioned studies show great potential across a number of applications for the usage of miniaturized NIRS. Such applications include, but are not limited to, identifying counterfeit products (\eg medicines and liquors), detecting allergens, or estimating the maturity of foods~\cite{nirsystems2002guide}. However, it is still time-consuming for developers to design and deploy applications using miniaturized NIRS, due to non-trivial programming tasks and lack of professional knowledge on NIRS. In fact, applications in existing studies show a major part of common components, including a mobile application and a remote server. The mobile application interacts with end-users and communicates with the miniaturized NIRS scanner and the remote server. The server processes the uploaded spectra from the client and returns interpreted information to the mobile application. The most important information is the classification results, as based on a machine learning model fed by the spectra. With these observations, we build and present \systemname, a ubiquitous NIRS sensing framework that leverages rapid development and deployment of miniaturized NIRS applications. The contributions of {\systemname} are 

\begin{itemize}\compresslist
    \item A generic server and a mobile application for developing general NIRS-based applications, as shown in Figure~\ref{fig:app_main} and \ref{fig:dashboard}. This contribution reduces the workload for development and deployment of applications using miniaturized NIRS for developers. As an implementation, we developed the framework using Python Django and the Android platform. 
    \item Generic NIRS spectrum processing methods on a server, reducing the requirement for professional knowledge of NIRS. 
    \item A straightforward logging-feedback method to collect NIRS spectra from the end-users to enrich the knowledge-base of miniaturized NIRS. The method can be used to crowdsource tasks for NIRS spectra collection.
\end{itemize}
\systemname~is open sourced in \cite{ubinirs}.

\begin{figure*}[t!]
    \centering
    \includegraphics[width=.95\textwidth]{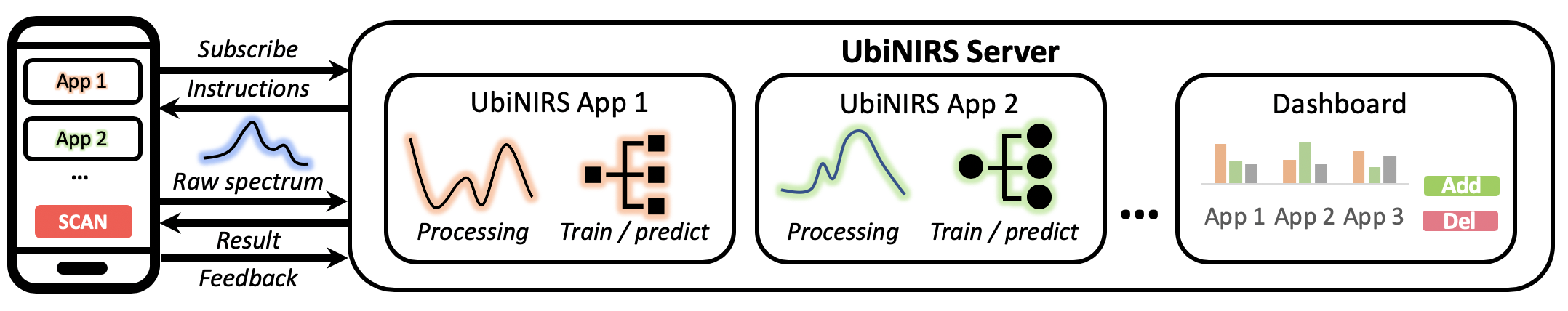}
    \caption{System Overview for \systemname.}
    \label{fig:system_overview}
    \vspace{-0.5em}
\end{figure*}

\section{\systemname~Framework}
\begin{marginfigure}[-34pc]
    \begin{minipage}{\marginparwidth}
        \centering
        \includegraphics[width=1.0\marginparwidth]{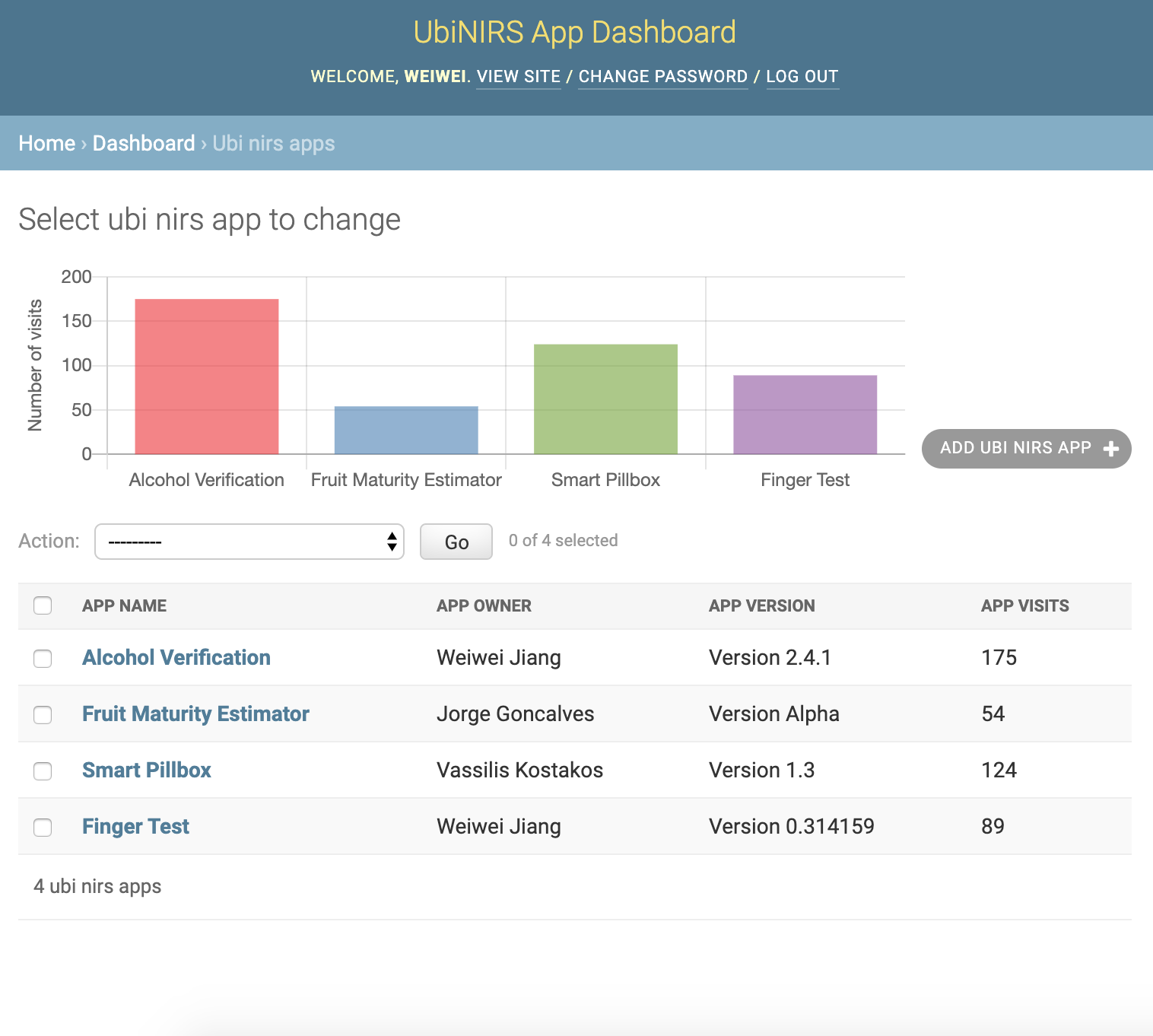}
        \caption{\systemname~dashboard.} 
        \label{fig:dashboard}
    \end{minipage}
\end{marginfigure}

\begin{marginfigure}[-14pc]
  \begin{minipage}{\marginparwidth}
    \centering
    \includegraphics[width=1.0\marginparwidth]{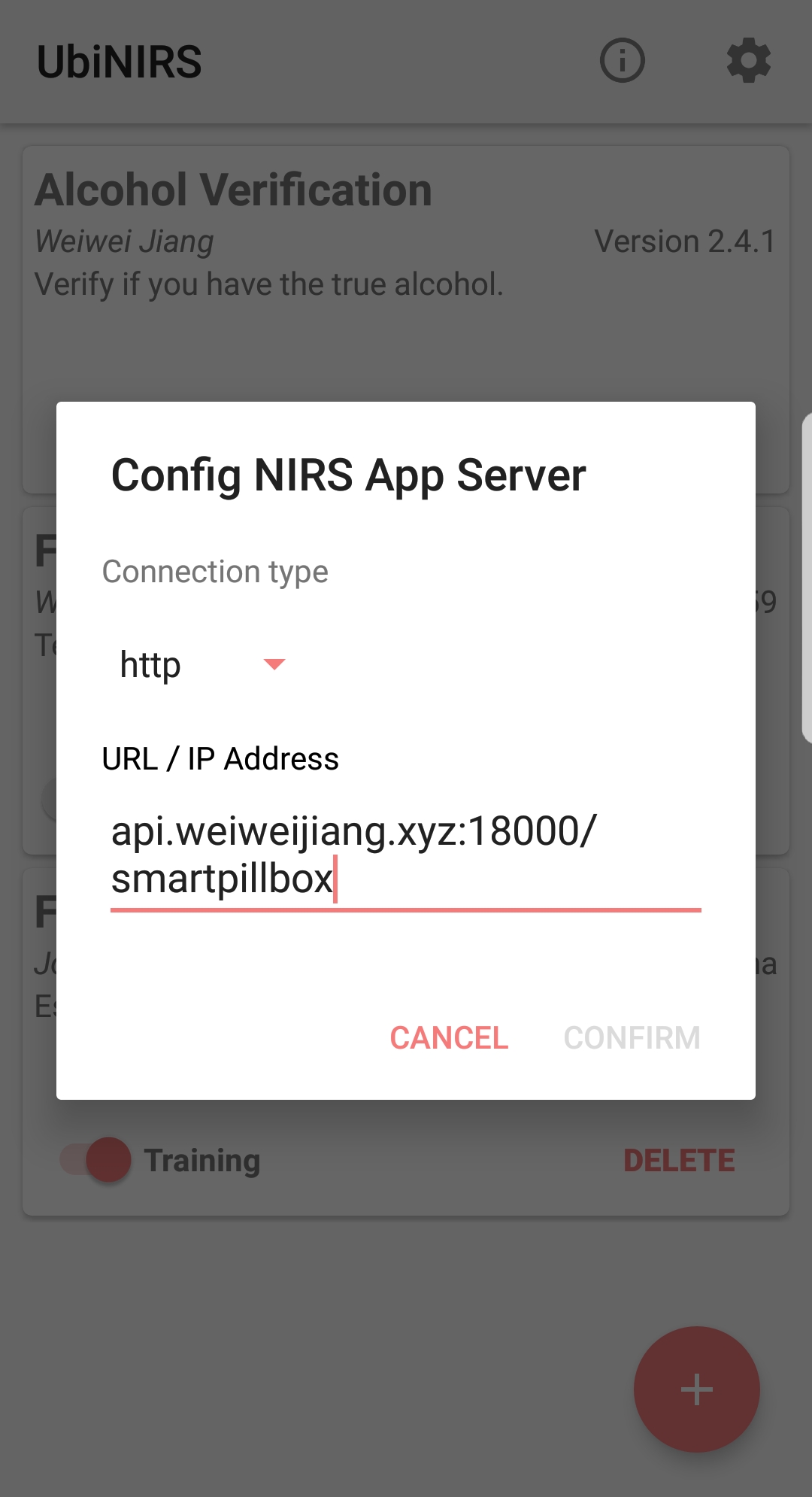}
    \caption{Subscribing a {\systemname} instance}~\label{fig:mobile_app}
  \end{minipage}
\end{marginfigure}

Figure~\ref{fig:system_overview} presents a system overview of \systemname. There are only two steps to start a {\systemname} instance: 1) Creating a {\systemname} instance in the dashboard using a web browser, as shown in Figure~\ref{fig:dashboard}, which automatically generates a URL to access the application. 2) The end-user registers the created {\systemname} instance in the mobile application using the generated URL, as shown in Figure~\ref{fig:mobile_app}. If necessary, the developer can further customize the server for additional functions, such as updating the spectrum processing methods or the machine learning methods, without any modifications to the mobile application.

\subsection{Session procedure}
Each usage session, \ie, the end-user scans an object and acquires the result, includes the following steps:
\begin{enumerate}\compresslist \vspace{-0.5em}
    \item \textbf{Start instance}: The end-user chooses a registered {\systemname} instance in the mobile application to send a starting request to the server. The server replies with instructions shown to the end-user for further actions. For example, when scanning pills, the end-user should put the flattest surface without carvings on the scanning window, then select ``scan''. The user can choose to either train the machine learning model in the server, or identify objects using the pre-train model. 
    \item \textbf{Scan object}: The end-user follows the instructions and pushes the ``scan'' button. The mobile application then sends a command to the NIRS scanner to perform a scan and retrieve a raw NIRS spectrum via Bluetooth. The spectrum is uploaded to the server for analysis. 
    \item \textbf{Receive result}: The server processes the raw spectrum and runs the machine learning model for either training the model or making an estimation using the processed spectrum. The result is sent back to the client.
    \item \textbf{Give feedback}: As an option, the end-user can send feedback for this session to the server for crowdsourcing tasks, such as validating the results, etc.  
\end{enumerate} \vspace{-0.5em}

% \marginpar{%
%   \vspace{-200pt} \fbox{%
%     \begin{minipage}{0.925\marginparwidth}
%         \textbf{Session Procedure} \\
%         \vspace{1pc} 
        
%     \end{minipage}}\label{sec:sidebar} 
% }

\subsection{Model Training}
\begin{marginfigure}[0pc]
  \begin{minipage}{\marginparwidth}
    \centering
    \includegraphics[width=0.9\marginparwidth]{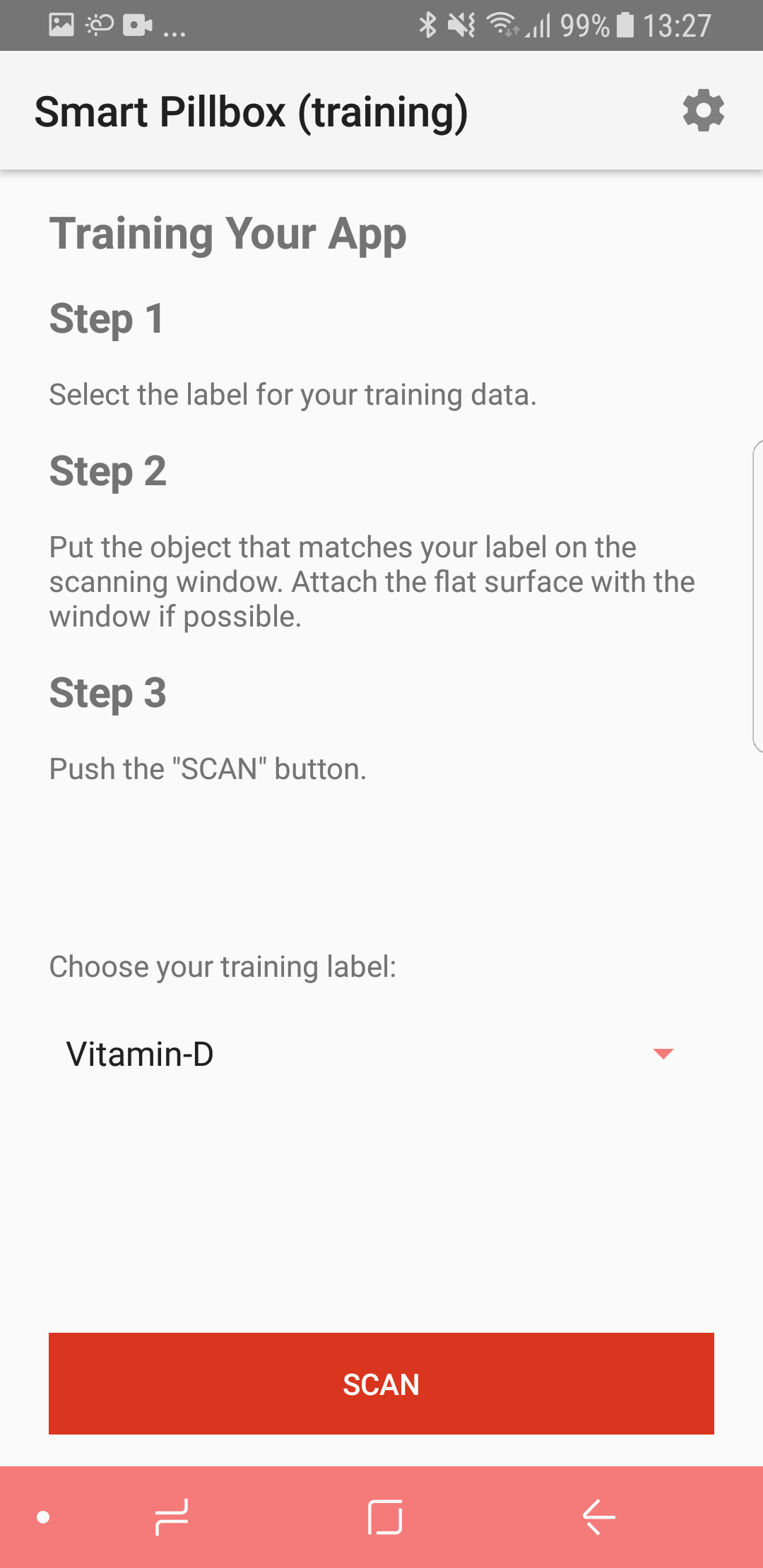}
    \caption{Instructions for training.}
    \label{fig:app_train}
  \end{minipage}
\end{marginfigure}

When a {\systemname} instance is created, the developer can upload reference spectra to train the machine learning model. As shown in Figure~\ref{fig:app_train}, the end-user can also train the model with the uploaded spectra for labeled samples \textit{in situ} as described above. 

% \begin{marginfigure}[-0pc]
%   \begin{minipage}{\marginparwidth}
%     \centering
%     \includegraphics[width=0.9\marginparwidth]{figures/cats}
%     \caption{Figure for App log}~\label{fig:crowdsourcing}
%   \end{minipage}
% \end{marginfigure}
% \vspace{1em}

\subsection{Spectrum Crowdsourcing}
An additional challenge for the use of miniaturized NIRS is constructing a knowledge-base for various objects. According to the literature, it is recommended to have at least $12$ spectra for each object in the knowledge-base~\cite{nirsystems2002guide}. Collecting these spectra requires increasing efforts with larger knowledge-base. Furthermore, some samples may not be directly accessible to the developer, obstructing inclusion in the knowledge-base. 

Our \systemname~framework can be used to address the aforementioned challenge by collecting information through crowd-sourcing tasks. End-users can train the model as described above, or provide feedback to the result returned by the trained model. The uploaded spectra are stored on the server and can be used for online training or validation.

\subsection{Implementation}
We have implemented the \systemname~platform using the Django~2 library~\cite{django2} in Python~3. The spectrum processing methods and machine learning models are based on the `\textit{scipy}' and `\textit{sklearn}' packages \cite{scipy}, as adopted from existing literature~\cite{jiang2019probing, klakegg2018assisted}. The mobile application is implemented for Android and uses the KS Technologies NIRScan Nano Android library~\cite{nirscanandroid}.

% \begin{figure}
%   \includegraphics[width=0.9\columnwidth]{figures/sigchi-logo}
%   \caption{Insert a caption below each figure.}~\label{fig:sample}
% \end{figure}

% \begin{table}
%   \centering
%   \begin{tabular}{l r r r}
%     % \toprule
%     & & \multicolumn{2}{c}{\small{\textbf{Test Conditions}}} \\
%     \cmidrule(r){3-4}
%     {\small\textit{Name}}
%     & {\small \textit{First}}
%       & {\small \textit{Second}}
%     & {\small \textit{Final}} \\
%     \midrule
%     Marsden & 223.0 & 44 & 432,321 \\
%     Nass & 22.2 & 16 & 234,333 \\
%     Borriello & 22.9 & 11 & 93,123 \\
%     Karat & 34.9 & 2200 & 103,322 \\
%     % \bottomrule
%   \end{tabular}
%   \caption{Table captions should be placed below the table. We
%     recommend table lines be 1 point, 25\% black. Minimize use of
%     table grid lines.}~\label{tab:table1}
% \end{table}

\section{Conclusion}
In conclusion, we present \systemname, a software framework for miniaturized NIRS-based applications. Our system greatly reduces the bar of developing NIRS-based applications, enables rapid deployment of material-level object identification application for miniaturized NIRS non-expert end-users. 

\bibliographystyle{abbrv}
\bibliography{main}

\balance{} 

\end{document}